\documentclass[aps,pra,eqsecnum,superscriptaddress,showpacs,showkeys,11pt]{revtex4}
\usepackage{epsfig}
\usepackage{graphicx}

\usepackage{amsmath}

\usepackage{bm}

\newcommand{\tr}{\mathop{\text{Tr}}\nolimits}
\newcommand{\ket}[1]{|{#1}\rangle}
\newcommand{\bra}[1]{\langle{#1}|}

\newcommand{\beq}{\begin{equation}}
\newcommand{\eeq}{\end{equation}}
\newcommand{\barr}{\begin{eqnarray}}
\newcommand{\earr}{\end{eqnarray}}

\begin{document}

\title{Robustness of optimal working points for non-adiabatic
holonomic quantum computation }

\author{Antonio Trullo} \affiliation{Dipartimento di Fisica,
Universit\`a di Bari,
        I-70126  Bari, Italy}
\author{Paolo Facchi}
\affiliation{Dipartimento di Matematica, Universit\`a di Bari,
        I-70125  Bari, Italy}
\affiliation{INFN, Sezione di Bari, I-70126 Bari, Italy}
\author{Rosario Fazio} \affiliation{NEST-CNR-INFM and
Scuola Normale Superiore, I-56126 Pisa, Italy}
\affiliation{International School
for Advanced Studies (SISSA), I-34014 Trieste, Italy}
\author{Giuseppe~Florio} \affiliation{Dipartimento di Fisica,
Universit\`a di Bari,
        I-70126  Bari, Italy}
\affiliation{INFN, Sezione di Bari, I-70126 Bari, Italy}
\author{Vittorio Giovannetti} \affiliation{NEST-CNR-INFM and
Scuola Normale Superiore, I-56126 Pisa, Italy}
\author{Saverio Pascazio} \affiliation{Dipartimento di Fisica,
Universit\`a di Bari,
        I-70126  Bari, Italy}
\affiliation{INFN, Sezione di Bari, I-70126 Bari, Italy}

\date{\today}

\begin{abstract}
Geometric phases are an interesting resource for quantum
computation, also in view of their robustness against decoherence
effects. We study here the effects of the environment on a class of
one-qubit holonomic gates that have been recently shown to be
characterized by ``optimal" working times. We numerically analyze
the behavior of these optimal points and focus on their robustness
against noise.
\end{abstract}



\maketitle

\section{Introduction}
\label{sec:Introduction}

Quantum algorithms based on geometric phases
\cite{shapere,bohm} are attracting increasing interest in
quantum computation \cite{nielsen,casati}. The related quantum gates
representing the unitary transformations on a register of qubits do
not have a dynamical origin: the Hamiltonian depends on time through
a set of control parameters that change by following suitable closed
loops in the associated parameter space; in the adiabatic limit the
dynamical contribution to the evolution can be factorized and the
features of the quantum gate depend only on the topological
structure of the manifold.

Geometric quantum computation has been investigated using both
Abelian~\cite{jones} and non-Abelian \cite{zanardi} holonomies.
There have been several proposals for their implementation using ion
traps \cite{duan}, Josephson junctions \cite{falci,faoro} and
semiconductors \cite{solinas1}. Since all physical devices interact
with their environment, one must carefully analyze the onset of
decoherence \cite{giulini} and its detrimental effects against
realistic implementations of quantum gates and algorithms. In
particular, the effects of noise for non-Abelian holonomies in open
quantum systems have been recently investigated in a series of
articles \cite{solinas2,fuentes,florio,parodi,sarandy2}. In
\cite{florio} we studied a class of one-qubit gates
implemented on a four-level (``tripod") system \cite{duan}, focusing
on non-adiabatic effects and bringing to light the presence of
fidelity revivals, namely an infinite number of (optimal) times at
which the fidelity reaches unity.

In this article we shall investigate the behavior of the fidelity at
the first of these optimal working points and study its robustness
against noise effects. The deviations from the ideal (noiseless)
case will be numerically analyzed as a function of the strength of
the noise (the coupling of the system with its environment) and a
heuristic definition on robustness will be introduced.

This paper is organized as follows. We review the concept of
holonomy in Section \ref{sec: Non-Abelian Geometric Phases} and
briefly introduce the specific tripod system \cite{duan} in
Section \ref{sec: Free Ideal Evolution}, where we focus on the
role of non-adiabatic effects. In Section \ref{sec: Noise and
Master Equation} we outline the main features of the master
equation for time dependent Hamiltonians: this is numerically
solved in Section~\ref{sec:results} in order to analyze the
behavior of the optimal working points in the presence of noise.
In Section~\ref{sec:conclusions} we and discuss the robustness of
our gates.

\section{Abelian and non abelian holonomies}
\label{sec: Non-Abelian Geometric Phases}

We consider a system governed by a non degenerate Hamiltonian that
depends on time through a set of parameters, adiabatically covering
a closed loop in the parameter space. Under these conditions, the
final state exhibits, in addition to the dynamical phase, also a
geometric phase, whose structure depends only on the topological
properties of the parameter manifold \cite{berry}. If the
Hamiltonian has some degeneracies, a loop in the parameter space
involves more complex geometric transformations
\cite{wilczeck}. We suppose that the family of Hamiltonians $H(x(t))$
($x^{\mu}(t)$ being a set of parameters) is iso-degenerate, i.e.\
that the dimensions of its eigenspaces do not depend on the
parameters and the eigenprojections $P_m(x(t))$ ($m$ denoting the
eigenvalue) have a smooth dependence on $t$ (at least twice
continuously differentiable). In particular, this entails the
absence of level crossing between different eigenspaces. $H(t)$ can
be decomposed by using its instantaneous eigenprojections
$H(t)=\sum_{m}\epsilon_m(t) P_m(t)$, with
$P_m(t)=\sum_k|m_k(t)\rangle\langle m_{k}(t)|$ and $k$ the
degeneracy index. We define the operator $R$, that transports every
eigenprojection from $t_0$ to $t$, and its hermitian generator
$D(t,t_0)$,
\beq R(t,t_0) P_m(t_0) = P_m(t) R(t,t_0),\,\,\qquad
D(t,t_0)= -i R(t,t_0)^\dagger
\frac{\partial}{\partial t}R(t,t_0).
\label{eq:intertwin}
\eeq
In the adiabatic limit the evolution of the state remains confined
in the degenerate eigenspaces and the evolution operator $U$ becomes
block-diagonal. In the case of cyclic evolution ($P_m(t)=P_m(t_0)$)
\begin{equation}\label{eq:Uadiab}
U(t,t_0)\sim \sum_m P_m(t_0)e^{-i  \int_{t_0}^{t} \epsilon_m(s)
ds} U_{\mathrm{ad}}^m P_m(t_0),\qquad U_{\mathrm{ad}}^m
=\textbf{P}\exp\left\{-\oint_C A^m(x)\right\},
\end{equation}
and the geometric evolution is given by a path ordered integral
($\textbf{P}$ in the above formula) of the adiabatic connection
$A^m(x)=\sum_\mu A^m_\mu dx^\mu$, with
\barr A^m_\mu(x(t)) = P_m(x(t_0)) R^{\dag}(x(t),x(t_0))
\frac{\partial}{\partial x^\mu}R(x(t),x(t_0))P_m(x(t_0)).
\label{eq: adiabatic connection} \earr
If the eigenvalues $\epsilon_m$ are time-independent and the
connection $D$ is piecewise constant (i.e.\ $D(t,t_0)=D(t_0,t_0)$
$\,\,\forall s\,\in\,[t,t_0]$) the evolution operator reduces to the
useful expression \cite{florio}
\beq\label{eq:UDt}
U(t,t_0)=e^{i(t-t_0)D(t_0,t_0)}\,e^{-i(t-t_0)(H(t_0)+D(t_0,t_0))}.
\eeq
We will study a large class of gates where the above hypoteses are
satisfied and one can exactly evaluate the time evolution, including
all non-adiabatic effects, by making use of (\ref{eq:UDt}).

\section{Free ideal evolution}
\label{sec: Free Ideal Evolution}

\begin{figure}
\begin{center}
\includegraphics[width=0.8\textwidth]{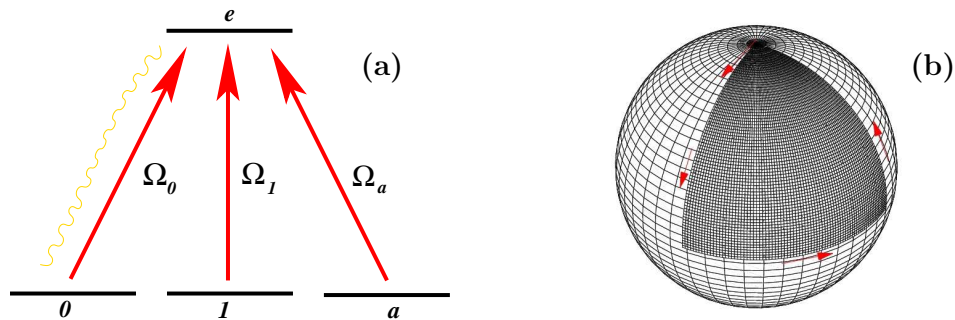}
\end{center}
\caption{(a): Scheme of a tripod system. 0 and 1 are the
computational levels, while $a$ is an ancilla state used for the
intermediate steps of the transformation. The three degenerate
levels are connected with an upper level $e$ by time dependent
Rabi frequencies $\Omega_j(t)$. The yellow wavy line represents
the noise, that induces additional transitions only between $0$
and $e$. (b): Path in parameter space for the realization of a NOT
gate. The solid angle spanned during the evolution is $\pi/2$.}
\label{fig: levels}
\end{figure}

We focus on the ``tripod" system introduced in \cite{duan} for
holonomic quantum computation: see Fig.\ \ref{fig: levels}(a), where
three degenerate levels are connected with a fourth one by Rabi
oscillations. The adiabatic evolution of this system was analyzed in
several articles for different experimental implementations
\cite{duan,faoro,solinas1}. Let us first review the ideal noiseless
case, taking into account also non-adiabatic effects. At time
$t=0$ the logical states $0$ and $1$ are encoded respectively in
the quantum states $|0\rangle$ and $|1\rangle$, while $|a\rangle$
is an ancilla state used as ``buffer'' during the evolution.
The Hamiltonian of the system reads $
    H(t)=|e\rangle(\Omega_0(t)\langle 0|+\Omega_1(t)\langle
    1|+\Omega_a(t)\langle a|)+ \mathrm{H.c.},
    $
where $\Omega_j(t)$ represent the time dependent Rabi frequencies of
the transitions. The loop in the parameter space is obtained by
varying $\Omega_j(t)$ ($j=0,1,a$). In our calculations we consider
$\Omega_j(t)\in \mathbf{R}, \forall t$. The eigenvalues of the
system are $
    \{0,\pm\sqrt{\Omega_0(t)^2+\Omega_1(t)^2+\Omega_a(t)^2}=\pm
    \Omega\}$,
where $0$ is 2-fold degenerate, corresponding to a 2-dimensional
(computational) eigenspace, and $\Omega$ is kept constant.
Therefore, the parameter space is the 2-sphere of radius $\Omega$,
$\{\Omega_j\in\mathbf{R}|\sum_j\Omega_j^2=\Omega^2\}$, shown in
Fig.\
\ref{fig: levels}(b). Introducing the parametrization
\begin{equation}
\Omega_1=\Omega\,\sin{\vartheta}\,\cos{\varphi}, \;
\Omega_0=\Omega\,\sin{\vartheta}\,\sin{\varphi}, \;
\Omega_a=\Omega\,\cos{\vartheta} , \label{eq: param trans} \eeq
the eigenstates take the form
\begin{eqnarray}
    |D_0(t)\rangle&=&
    \cos{\varphi}\,|0\rangle-\sin{\varphi}\,|1\rangle,  \nonumber\\
    |D_1(t)\rangle&=&\cos{\vartheta}\, \sin{\varphi}|0\rangle+\cos{\vartheta}\,
    \cos{\varphi}|1\rangle-\sin{\vartheta}|a\rangle, \label{eq: eigenstates in param}\\
    |D_\pm(t)\rangle&=&\big(\pm|e\rangle+\sin{\vartheta}\, \sin{\varphi}|0\rangle+
    \sin{\vartheta}\,
    \cos{\varphi}|1\rangle+\cos{\vartheta}|a\rangle\big)/\big(\sqrt{2}\big).\nonumber
\earr
The computational space
$C_S=\text{Span}\{|D_0(t)\rangle,|D_1(t)\rangle\}$ belongs to the
degenerate eigenvalue $0$, while $|D_\pm(t)\rangle$ are the bright
eigenstates belonging to $\pm\Omega$. One easily shows that for a
closed loop on the 2-sphere in the computational space the holonomy
(\ref{eq:Uadiab}) reads $U_{\mathrm{ad}}=\exp\left(i \sigma_y
\,\mathcal{\omega}\right)$, where
$\sigma_y=-i(\ket{D_0(t_0)}\bra{D_1(t_0)}-\ket{D_1(t_0)}\bra{D_0(t_0)})$
and $\omega$ is the solid angle enclosed by the loop. In particular,
if $\omega=\pi/2$, we have $U_{\pi/2}=\exp(i
\sigma_y \,\pi/2)= i \sigma_y$ (in the basis
$\{|D_0(t_0)\rangle,|D_1(t_0)\rangle\}$), that represents a NOT
transformation (up to a phase for the state $|D_0\rangle$).

Following the discussion of the previous section we discuss the
non-adiabatic corrections to this system. In order to use
Eq.~(\ref{eq:UDt}), we will consider the loop shown in Fig.\
\ref{fig: levels}(b), enclosing the solid angle $\pi/2$;
in the adiabatic limit (when $\Omega\,\tau \to \infty$,
$\tau$ being the total time of the cyclic evolution and $\Omega$ the
energy of the bright states) this path yields a NOT gate. The first
step consists in constructing the operator $D$ from Eq.~(\ref{eq:
eigenstates in param}) and the definition (\ref{eq:intertwin}). One
can see \cite{florio} that, as far as the rate of change of the
polar angles
is constant in each section of the path, $D(t,t_0)$ is piecewise
constant and we can use Eq.~(\ref{eq:UDt}) to evaluate the evolution
operator along the path shown in Fig.\
\ref{fig: levels}(b).

An interesting feature of the evolution is that it is factorized
in three terms. In the adiabatic limit it simplifies to
\begin{equation}
U_{\pi/2}(\Omega\tau)=U_{3}(\Omega\tau_3)\,U_{2}(\Omega\tau_2)\,U_{1}(\Omega\tau_1)
\stackrel{\tau \Omega\rightarrow\infty}{\longrightarrow}
    U_{\pi/2}^{\rm{ad}}(\Omega\tau)=\left(\begin{array}{cccc}0&1&0&0\\
    -1&0&0&0\\
    0&0&e^{-i\tau\Omega}&0\\
    0&0&0&e^{+i\tau\Omega}\\
    \end{array}\right) ,
\label{eq: U analytic in adiab lim}
\end{equation}
$\tau$ being the total evolution time needed for covering the loop
in the parameter space and $\tau_i=\alpha_i\tau$, with
$\sum_i\alpha_i=1$. This represents a NOT gate for the degenerate
subspace and yields (fast oscillating) dynamical phases for the
bright states.

In order to understand how close the evolution operator is to the
ideal one, we use the mean fidelity
\begin{eqnarray}
\langle F \rangle (\Omega\tau) &=& \frac{1}{4\pi}
\int d \cos \vartheta d \varphi F(\Omega \tau,\vartheta, \varphi)
\nonumber \\
&=& \frac{1}{4\pi}\int d \cos \vartheta d \varphi
\tr\{\sigma_{\rm{ad}}(\Omega\tau,\vartheta,\varphi)\sigma(\Omega\tau,\vartheta,\varphi)\},
\label{eq: def fidelity}
\end{eqnarray}
where
\begin{eqnarray}
\sigma(\Omega\tau,\vartheta,\varphi) &=&
U_{\pi/2}(\Omega\tau)\sigma(\vartheta,\varphi)U^{\dag}_{\pi/2}(\Omega\tau),
\label{eq:defsig} \\
\sigma_{\rm{ad}}(\Omega\tau,\vartheta,\varphi)&=&
U_{\pi/2}^{\rm{ad}}(\Omega\tau)\sigma(\vartheta,\varphi)U^{\rm{ad}\dag}_{\pi/2}(\Omega\tau)
,
\label{eq:defsigad}
\end{eqnarray}
$\sigma(\vartheta,\varphi)=|\vartheta,\varphi\rangle\langle\vartheta,\varphi|$
being the initial state (assumed to be pure). In practice, in our
analysis, $F$ will always be averaged over a finite set of input
states uniformly distributed on the Bloch sphere. The mean
fidelity is plotted in Fig.\ \ref{fig:ideal} as a function of the
adiabaticity parameter $\Omega\tau$.
\begin{figure}
\begin{center}
\includegraphics[width=0.6 \textwidth]{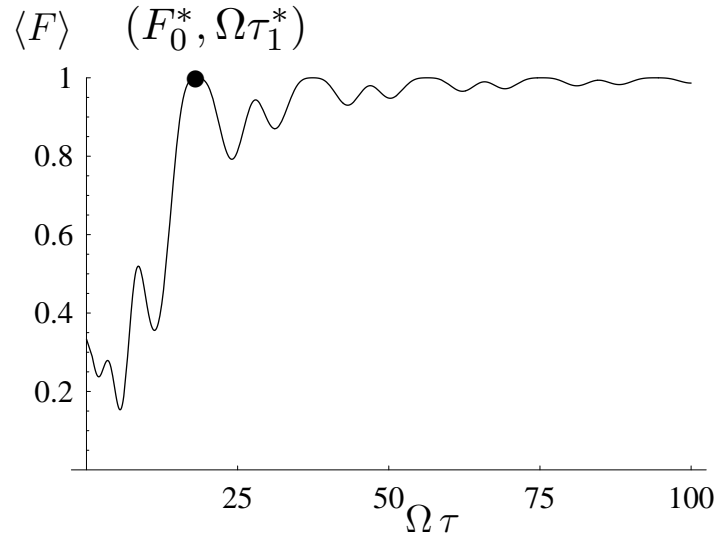}
\end{center}
\caption{Mean fidelity versus the cyclic time $\Omega \tau$
(noiseless case). $\Omega$ is the energy gap between the bright
and dark states. $\tau$ is the time needed to cover the loop shown
in Fig.~\ref{fig: levels}(b). The average is performed over a set
of initial states uniformly distributed on the Bloch sphere. The
dot on the first significant peak indicates the optimal working
point, $F_0^*=1$ and $\Omega\tau_1^*=18.25$.} \label{fig:ideal}
\end{figure}
Clearly, $\langle F \rangle$ asymptotically approaches unity (with
some oscillations), as expected (adiabatic limit). Notice that the
fidelity is exactly one for some \emph{finite} values of time,
$\tau=\tau^*_k$, that are independent of the initial state. In this
case the NOT transformation is perfect, even though one is far from
the adiabatic regime.

It is possible to show that, when the three arcs in the loop in
Fig.\ \ref{fig: levels}(b) are covered in equal times, one obtains
\beq
\label{eq:optimalNOT}
\tau^*_k=
\frac{3\pi}{2\Omega}\sqrt{16 k^2-1} \ , \qquad k\in \mathbf{N}^* .
\eeq
The first fidelity revival occurs for $k=1$
\beq\label{eq:optimal1}
\Omega\tau^*_1=\frac{3\pi}{2}\sqrt{15} = 18.25
\eeq
and is
indicated by a dot in Fig.\ \ref{fig:ideal}. These revivals (the
first one in particular) can be important for experimental
applications: in principle they would enable one to obtain a
perfect NOT transformation, without reaching the adiabatic regime.
It is important to notice that this result does not depend on the
initial state of the system but is a feature of the chosen path
(see \cite{florio} for details).

Finally, we emphasize that similar features (and in particular the
presence of the revivals in the non-adiabatic regime) hold for a
large class of gates. For transformations consisting in a loop that
starts at the pole and is composed of three geodesics given by two
arcs of meridians and an arc of the equator, enclosing a solid angle
$\omega=\pi/2n$ ($n\in \mathbf{N}^*$), there is a straightforward
generalization of Eq.~(\ref{eq:optimalNOT}): \beq
\label{eq: generalization of optimal times}
\tau^*_k(n)= \frac{(2n+1)\pi}{2n\Omega}\sqrt{16 k^2 n^2-1}. \eeq
This expression is valid provided that the loop is covered at a
constant angular speed [see Fig.\ \ref{fig: levels}(b)]:
$\dot{\vartheta}_{\rm arc 1}=\dot{\varphi}_{\rm arc
2}=\dot{\vartheta}_{\rm arc 3}=\mathrm{const}$. Reversing the
orientation of the loop leads to identical results.

\section{Master Equation for a Time-Dependent Hamiltonian}
\label{sec: Noise and Master Equation}

The interaction between a system and the environment is usually
analyzed in terms of a master equation. In the standard approach
to this problem one assumes that the Hamiltonian of the system is
time independent (see for instance \cite{gardiner}). For time
dependent Hamiltonians a slightly different approach is needed
\cite{davies,florio}. We consider a general Liouville operator
with a time dependent system Liouvillian \beq\label{eq:
liouvillianA} \mathcal{L}(t)=\mathcal{L}_0(t)+\lambda
\mathcal{L}_{SB}
=\mathcal{L}_S(t)\otimes1+1\otimes\mathcal{L}_{B}+\lambda
\mathcal{L}_{SB} , \eeq where $\lambda$ is the dimensionless
coupling constant representing the strength of the noise. The
evolution of density operator $\varrho(t)$, describing the system
and the environment, is governed by the von Neumann-Liouville
equation $\dot{\varrho}(t)=\mathcal{L}(t)\,\varrho(t)$. We assume
that there are no initial correlations between system and bath
(i.e.\ the initial state is factorized) and that the bath is in
equilibrium (e.g.\ in a thermal state). The main hypothesis in the
derivation of a master equation is that the typical timescale of
the evolution is much slower than the timescales characterizing
the bath. We shall also assume that the timescale related to the
rate of change of the system Hamiltonian is the slowest timescale
of our problem: this is clearly related to the adiabaticity of the
evolution. In other words, compared to the bath correlation time,
the evolution of $\mathcal{L}_S$ is always ``adiabatic." This is
assured by the condition \beq \label{eq: condition on LSA}
\tau_c\Delta\ll1 , \eeq where $\tau_c$ is the correlation time of
the bath and the energy gap $\Delta=
{\text{min}}\left|\epsilon_n(t)-\epsilon_m(t)\right|$
characterizes the rate of change of $\mathcal{L}_S$. Under these
conditions one gets
\begin{eqnarray}\label{eq:MEred}
\dot{\sigma}(t) =[\mathcal{L}_{S}(t)+\lambda^2\Gamma(t)]\sigma(t)
,
\end{eqnarray}
where $\sigma(t)= \tr_B\left\{\varrho(t)\right\}$ is the system
density matrix and $\Gamma(t)$ a time dependent dissipation
superoperator. Equation (\ref{eq:MEred}) is the same master equation
one would obtain by considering $\mathcal{L}_S(t)$ ``frozen" at time
$t$ and evaluating the decay rates and the frequency shifts at the
instantaneous eigenfrequencies
$\omega(t)=\epsilon_m(t)-\epsilon_n(t)$ of the system Liouvillian.

We consider now the physical system described in Sec.\ \ref{sec:
Free Ideal Evolution}. For simplicity let the environment affect
only the transitions between levels $|0\rangle$ and $|e\rangle$;
this is enough for our purposes. The total Hamiltonian is
$H_T(t)=H(t)+H_B+ \lambda H_{SB}$, where $H(t)$ is the system
Hamiltonian. The bath is bosonic, $H_B=\sum_k \omega_k{a_k}^\dag a_k
$ with $\omega_k$ the frequency of the $k$-th mode. The interaction
Hamiltonian is $H_{SB}=\sum_k \gamma_k (|0\rangle\langle e|
+|e\rangle\langle 0|)\otimes({a_k}^\dag+a_k)$, where $\gamma_k$ is
the coupling constant between the system and the $k$-th mode of the
bath. By using Eq.~(\ref{eq: eigenstates in param}) and the form of
the interaction Hamiltonian, we can obtain time dependent Lindblad
operators describing the transitions caused by the environment. In
the interaction picture generated by the operator $R$ defined in
(\ref{eq:intertwin}), the density operator
$\sigma_R(t)=R^\dag\sigma(t)R $ satisfies the following master
equation:
\begin{equation}\label{eq:MEforSystem}
\dot{\sigma}_R(t)=-i[H_S(0),\sigma_R(t)]-i[D(t,0),\sigma_R(t)]+\lambda^2
\Gamma(t)\sigma_R(t),
\end{equation}
where the Lamb shifts and the decay rates can be evaluated by
standard formulas \cite{florio,tasaki}, when one introduces the
appropriate thermal spectral densities.

\section{Fidelity and behavior of the optimal working point}
\label{sec:results}

Equation~(\ref{eq:MEforSystem}) was numerically integrated along the
loop in Fig.~\ref{fig: levels}(b) when the three arcs are covered at
a constant angular speed. The values of the Lamb shifts and decay
rates were assigned, somewhat arbitrarily, for illustrative
purposes. They correspond to a bath at very high temperature.

The behavior of the average fidelity (\ref{eq: def fidelity}), with
$\sigma(\Omega\tau,\vartheta,\varphi)$ numerically obtained from
(\ref{eq:MEforSystem}), is shown in Fig.\
\ref{fig:media}: from top to bottom, the dissipation constant
increases from $\lambda^2=0$ to 0.05. In the noiseless case (upmost
line) the fidelity tends to 1 when $\Omega\tau
\to \infty$ (adiabatic limit). This asymptotic value is not reached
monotonically: there are some oscillations, with maxima at $F=1$ in
the noiseless case. This is the case discussed in Section
\ref{sec: Free Ideal Evolution}: the NOT transformation is
perfect, even though one is far from the adiabatic regime, at the
time values given by (\ref{eq:optimalNOT}).

\begin{figure}
\begin{center}
\includegraphics[width=0.7\textwidth]{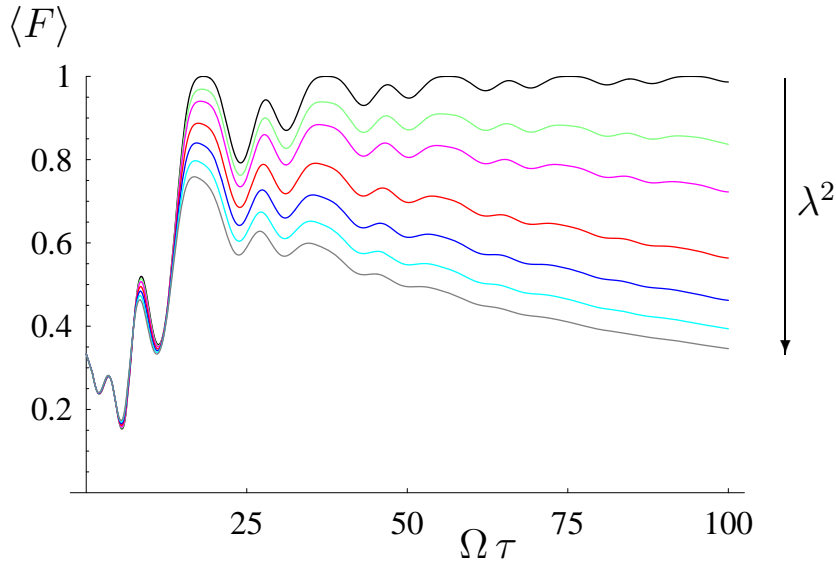}
\end{center}
\caption{Mean fidelity $\langle F \rangle$ versus cyclic time
$\Omega \tau$. The dissipation constant $\lambda^2$ increases from
top to bottom: $\lambda^2=0$ (noiseless case), 0.005, 0.01, 0.02,
0.03, 0.04 and 0.05. } \label{fig:media}
\end{figure}

Clearly, in the presence of noise, the fidelity decreases as the
time needed for the transformation increases. This can make it
difficult to obtain a pure geometrical transformation (because of
the necessary adiabatic condition). Therefore, it appears convenient
to take advantage of the presence of the peaks. As a matter of fact,
the fidelity decrease due to the noise is very small in the non
adiabatic regime and one can think of realizing the NOT gate by fine
tuning the total operation time. As the best performance is obtained
for the first peak of the fidelity (\emph{optimal operation point}),
we shall focus on the $\lambda$-dependence of the coordinates of the
first significant maximum, $F^*$ and $\tau^*$, and their deviation
from the noiseless values $F^*_0$ and $\tau^*_1$ (see Fig.\
\ref{fig:ideal}). It is important to stress that in the non
adiabatic regime the gate is no longer purely geometrical; in
principle it would be possible to extract the geometric
contribution, but one would not gain any additional information,
useful for experimental purposes.

A critical issue is the total amount of noise. In the simulations in
Fig.\ \ref{fig:media} we considered a noise strength $\lambda^2$
ranging from $0.005$ to $0.05$. However, a realistic physical
estimate, using thermal spectral densities, would yield a noise
level below $0.005$ \cite{florio}. In this regime the fidelity at
the optimal point reaches values greater than $0.9$. From this
result it is clear that we can exploit the optimal times for
realizing the NOT transformation with a relatively high fidelity
even in absence of additional control.

\begin{figure}
\includegraphics[width=0.99\textwidth]{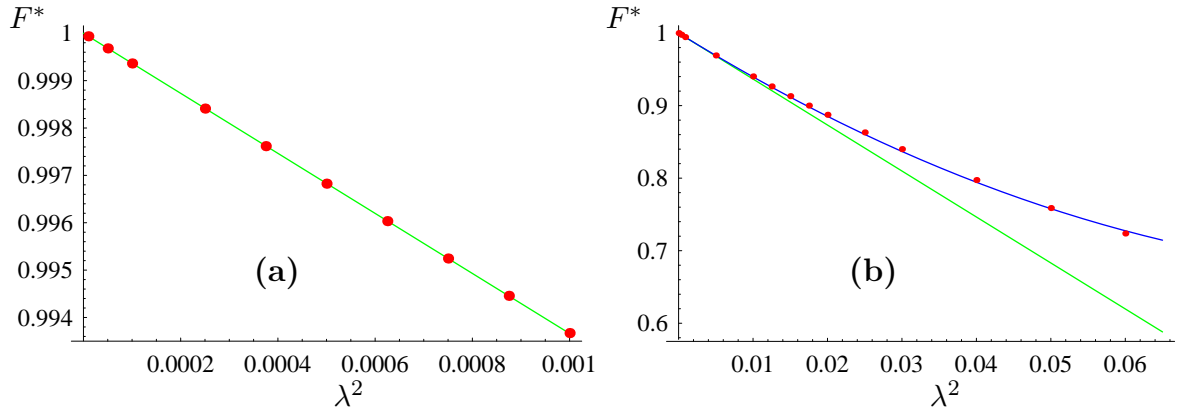}
\caption{Maximum value of the mean fidelity $F^*$ vs noise
(coupling to the bath) $\lambda^2$. (a) Small coupling: the fit
yields $F^* = 1 - 6.34 \lambda^2$. (b) Larger coupling: the fit
yields $F^* = 1 - 6.34 \lambda^2 + 29.93 \lambda^4$; linear fit as
in (a). The error bars are always smaller than the size of the
points. }
\label{graficoflambdagran1}
\end{figure}

It is important to understand how the optimal time and the
corresponding value of fidelity change by increasing the strength of
the noise. This should yield information about the robustness of
holonomic quantum computation against the detrimental effects of
noise. Figure \ref{graficoflambdagran1}(a) shows the behavior of
$F^*$ for small noise: the points are the result of a numerical
analysis and the continuous line is the fit
\beq
\label{eq:Fvslabda} F^* = 1 - F_2 \lambda^2,
\eeq
yielding $F_2=6.34$. The agrement is excellent and enables one to
conclude that fidelity decreases linearly with $\lambda^2$ for
$\lambda^2 \leq 10^{-3}$. The behavior of $F^*$ for larger values
of $\lambda^2$ is displayed in Fig.\ \ref{graficoflambdagran1}(b).
The fit is \beq \label{eq:Fvslabdalarge} F^* = 1 - F_2 \lambda^2 +
F_4 \lambda^4, \eeq with $F_4=29.93$. Observe that $\lambda^2
\simeq 6\times10^{-2}$ is a very large (unphysical) value. The
conclusion that fidelity decreases as $\lambda^2$ (for small
$\lambda$) is to be expected from a perturbation expansion of the
master equation (\ref{eq:MEforSystem}).

Let us now analyze the behavior of the optimal time $\tau^*$ vs
$\lambda^2$. Figure \ref{graficotaulambdagran1}(a) displays the
otpimal time vs $\lambda^2$ in the small coupling case. The values
of $\tau^{*}$ are more affected by errors due to the approximations
introduced by the numerical calculations. The optimal time decreases
by increasing $\lambda^2$. The fit yields
\beq \label{eq:tauvslabda}
\tau^* = \tau^*_1 - \tau_2 \lambda^2,
\eeq
 with $\tau_2 = 59.40 \Omega^{-1}$, the value $\tau_1^*=18.25\Omega^{-1}$ being obtained
analytically from Eq.\ (\ref{eq:optimal1}). Again, the linear fit
is in good agreement with the data and enables one to conclude
that the optimal time decreases linearly with $\lambda^2$ for
$\lambda^2 \leq 10^{-3}$. The behavior of $\tau^*$ for larger
values of $\lambda^2$ is displayed in Fig.\
\ref{graficotaulambdagran1}(b). In this case the fit is
\beq
\label{eq:tauvslabdalarge} \tau^* = \tau^*_1 - \tau_2 \lambda^2 +
\tau_4 \lambda^4 - \tau_6 \lambda^6,
\eeq
with $\tau_4=990.65\Omega^{-1}$ and $\tau_6=7655.95\Omega^{-1}$.

\begin{figure}
\includegraphics[width=0.99\textwidth]{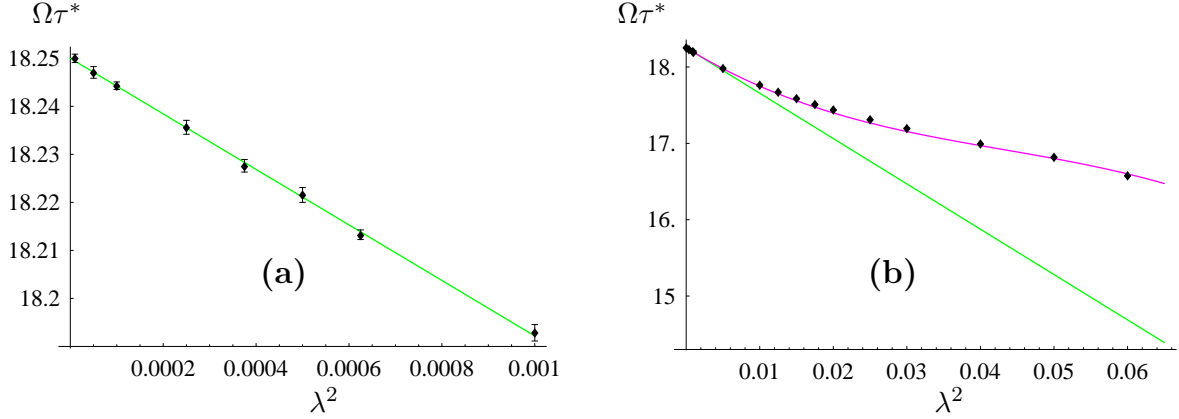}
\caption{Optimal time $\tau^*$ vs noise (coupling to the bath)
$\lambda^2$. The values of $\tau^{*}$ are more affected by error
than those of $F^*$. (a) Small coupling: the fit yields
$\Omega\tau^* = \Omega\tau^*_1 - 59.40 \lambda^2$, with
$\Omega\tau^*_1=18.25$, the theoretical value (\ref{eq:optimal1}).
(b) Larger coupling: the fit yields $\Omega\tau^* = \Omega\tau^*_1
-59.40 \lambda^2 +990.65 \lambda^4-7655.95 \lambda^6$; linear fit
as in (a). }
\label{graficotaulambdagran1}
\end{figure}

The behavior of the point $(F^*,\tau^{*})$ is shown in Fig.\
\ref{graficoftaugran1}, both for small (a) and larger (b) values
of $\lambda^2$. From Eqs.\ (\ref{eq:Fvslabda}) and
(\ref{eq:tauvslabda}) one obtains
\beq
F^{*}=1-\frac{F_2}{\Omega\tau_2}\Omega(\tau^*_1-\tau^*)=1 +
0.11\Omega (\tau^*-\tau^{*}_{1}) .
\label{eq:ftaulinear}
\eeq
This is the small coupling situation displayed in Fig.\
\ref{graficoftaugran1}(a). Notice that the mean fidelity increases
linearly with $\tau^{*}$ in this regime. In the presence of noise,
optimal quantum gates are less precise and slightly faster than
the ideal ones. A further analysis of this dependence for larger
values of the noise [using Eqs.\ (\ref{eq:Fvslabdalarge}) and
(\ref{eq:tauvslabdalarge})] yields an involved algebraic
expression that includes higher order corrections. The behavior is
shown in Fig. \ref{graficoftaugran1}(b). The evolution of the
optimal working point is summarized in Fig.\ \ref{fig:mediabis}.

\begin{figure}
\includegraphics[width=0.99\textwidth]{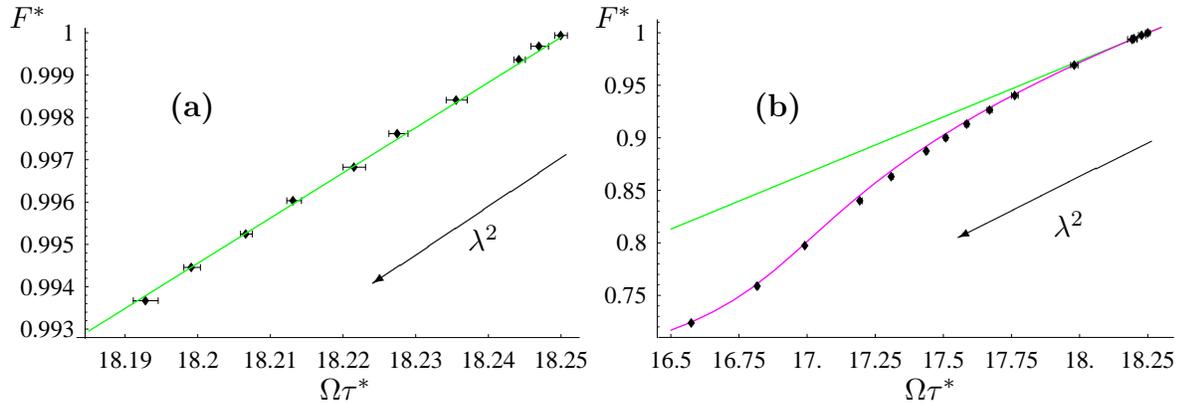}
\caption{Maximum value of the mean fidelity vs the optimal working
time $\tau^*$. (a) Small coupling: the fit yields $F^* = 1 + 0.11
\Omega (\tau^*-\tau^{*}_{1})$. (b) Larger coupling; the errors are
included in the size of the points and the linear fit is as in
(a).}
\label{graficoftaugran1}
\end{figure}

\begin{figure}
\begin{center}
\includegraphics[width=0.7\textwidth]{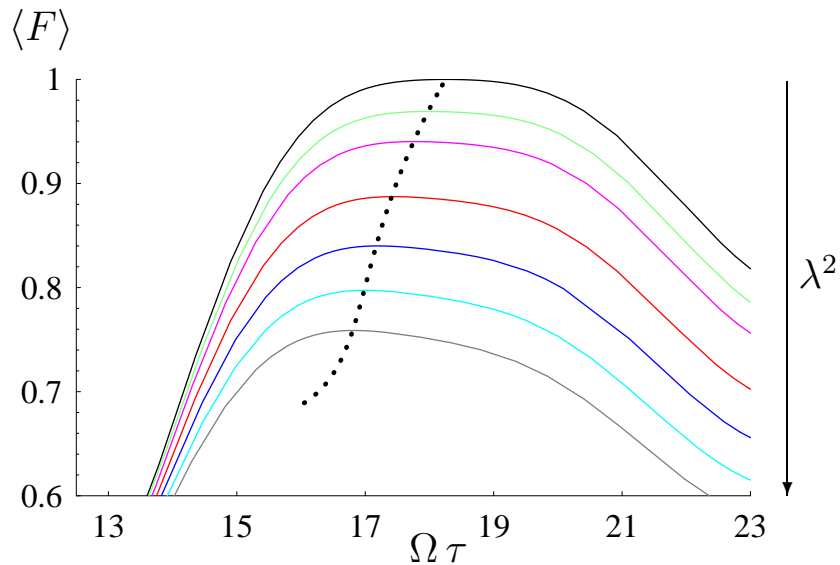}
\end{center}
\caption{Mean fidelity $\langle F \rangle$ versus cyclic time
$\Omega \tau$: a closer look at the evolution of the optimal point
in Fig.\ \ref{fig:media}. The dissipation constant $\lambda^2$
increases from top to bottom: $\lambda^2=0$ (noiseless case),
0.005, 0.01, 0.02, 0.03, 0.04 and 0.05. The dotted line is the fit
described in the text.}
\label{fig:mediabis}
\end{figure}

\section{Robustness and concluding remarks}
\label{sec:conclusions}

In order to shed some light on the robustness of the quantum gate
in the neighborhood of the optimal working point it is useful to
compare the optimal fidelity $F^*$ with the fidelity obtained in
the adiabatic limit $F_{\rm{adiab}}$. Let us observe that in
general, the analysis of decoherence in geometric computation
raises a critical issue in connection with adiabatic evolutions,
that cannot be too slow, as decoherence would eventually destroy
any interference. This inevitably introduces an element of
arbitrariness in the following definitions.

We shall evaluate the performance of the optimal (non adiabatic)
quantum gate, as compared to its adiabatic limit, throught a
``robustness'' parameter:
\beq
\textit{R}=\frac{F^*-F_{\rm{adiab}}}{F^*}.
\eeq
By glancing at Figs.\  \ref{fig:ideal} and \ref{fig:media} it is
apparent that the adiabatic limit is practically attained already at
the third peak, namely for $\tau = \tau^*_3$ in
Eq.~(\ref{eq:optimalNOT}). We therefore take
\beq
F_{\rm{adiab}}\simeq
F(\Omega\tau^*_{3})=F\left(\frac{3\pi}{2}\sqrt{143}\right).
\eeq
The dependence of $R$ on $\lambda^2$ is displayed in Fig.\
\ref{fig:robparameter}. Clearly, the (relative) robustness of the
optimal gate is larger for larger noise levels. We notice the
presence of a linear regime for small coupling.
\begin{figure}
\includegraphics[width=0.6\textwidth]{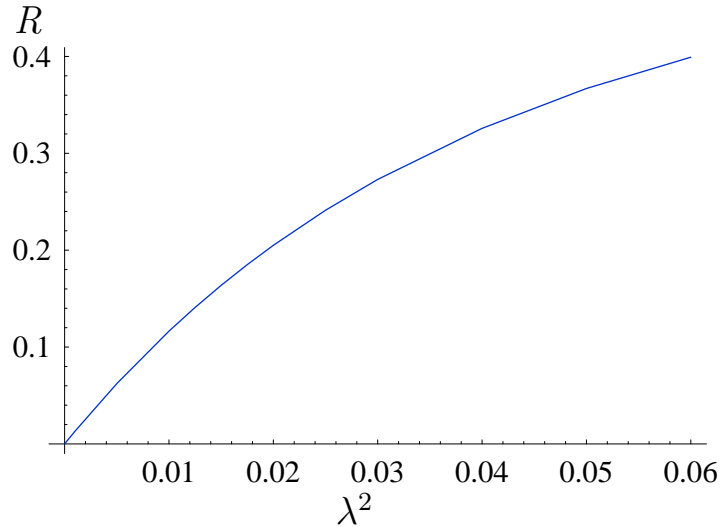}
\caption{Robustness parameter $R$ vs
$\lambda^2$.}\label{fig:robparameter}
\end{figure}

Although we focused our attention on the particular physical system
shown in Fig.\ \ref{fig: levels}(a), our conclusions are rather
general and are valid for other physically relevant situations.
There exist some values of the duration of the evolution for which
the fidelity is 1, even though one is far from the adiabatic regime.
As already emphasized at the end of Sec.\ \ref{sec: Free Ideal
Evolution}, these results can be extended to more general loops,
yielding optimal times like in Eq.~(\ref{eq: generalization of
optimal times}).

The presence of these optimal peaks is important for experimental
applications: if the total operation time can be tuned to the first
peak, one can realize a transformation that is the most robust
against noise. Moreover the maximum value of the fidelity exhibits
different regimes: for small coupling it decreases linearly with
$\lambda^2$. In general, the gate can be considered robust when
compared to the standard adiabatic one. This can be of interest for
experimental applications, if one aims at introducing further
control. The case of two-qubit gates is not trivial and is at
present under investigation.

\acknowledgments
This work was partly supported by the European Community through the
Integrated Project EuroSQIP and by the bilateral Italian– Japanese
Projects II04C1AF4E on "Quantum Information, Computation and
Communication" of the Italian Ministry of Instruction, University
and Research.


\end{document}